\documentclass[aps,showpacs,preprintnumbers,amsmath,amssymb]{revtex4}
\UseRawInputEncoding
\usepackage{dcolumn}

\usepackage{color}
\usepackage{float}
\usepackage{graphics,epsfig}
\usepackage{epstopdf}
\usepackage{graphicx}
\usepackage{multirow}
\usepackage{dcolumn}
\usepackage{bm}
\usepackage{gensymb}

\begin{document}
\baselineskip=0.8 cm
\title{\bf Kerr black hole shadows cast by extraordinary light rays with Weyl corrections}
\author{
Songbai Chen$^{1,2}$\footnote{Corresponding author: csb3752@hunnu.edu.cn},
Jiliang Jing$^{1,2}$ \footnote{jljing@hunnu.edu.cn}}
\affiliation{ $ ^1$ $^1$Department of Physics, Institute of Interdisciplinary Studies, Key Laboratory of Low Dimensional Quantum Structures
    and Quantum Control of Ministry of Education, Synergetic Innovation Center for Quantum Effects and Applications, Hunan
    Normal University,  Changsha, Hunan 410081, People's Republic of China
    \\
    $ ^2$Center for Gravitation and Cosmology, College of Physical Science and Technology, Yangzhou University, Yangzhou 225009, People's Republic of China}

\begin{abstract}
\baselineskip=0.6 cm
\begin{center}
{\bf Abstract}
\end{center}

We investigate the equation of motion for photons with Weyl corrections in a Kerr
black hole spacetime in a small coupling case. Our results show that Weyl corrections yield phenomena
of birefringence. The light rays propagating in the spacetime are separated into the ordinary rays and the extraordinary rays,  and the propagation of the latter depends on the corrections. We probe the effects of Weyl corrections on the Kerr black hole shadows casted by the extraordinary rays and find that such corrections result in a weak stretching or squeezing in the vertical direction for the black hole shadows. Finally, we also study the change of the length of the Near-Horizon Extremal Kerr line (NHEK line) with Weyl corrections. These features could help us to understand the electrodynamics with Weyl corrections from black hole shadows.

\vspace{0.2cm}

{\bf Black hole shadow, Weyl correction, birefringence, extraordinary rays}

\end{abstract}

 \maketitle
\newpage
\section{Introduction}

The images of the supermassive black holes M87* \cite{fbhs1,fbhs6,fbhs12,fbhs13} and Sgr A* \cite{fbhs1222,fbhs17} open a new era of testing gravity in strong field regimes. The brightness and polarization patterns of surrounding emission regions in black hole images carry a wealth of information about electromagnetic emissions \cite{kerr1,kerr2,bhs0,bhs1,bhs2,bhs3,bhsp1}, which provides a powerful method to probe the electromagnetic interaction, matter distribution, and accretion process in the vicinity of black holes. In general, black hole images depend on the parameters of the background black hole, the dynamical properties of the photon, and the interactions between the photon and other fields.

Most current studies of black hole images are based on the standard Einstein-Maxwell theory in which there is only a quadratic term of Maxwell tensor directly related to electromagnetic field \cite{sw,swo,astro,chaotic,binary,sha18,my,swo7,swo8,swo9,swo10,swo11,rs,dark1,dark2,dark3,dark4,min22}. Recently, black hole images have been investigated in some interesting extensions of the Einstein-Maxwell theory that contained more interactions with the electromagnetic field. These extra interactions modify the Maxwell equations and change the paths of photons moving in spacetimes, which inevitably affect the sizes and shapes of black hole shadows. Refs. \cite{qed1,qed2,qed3} showed that the quantum electrodynamic corrections from the Euler-Heisenberg effective Lagrangian yields a birefringence phenomenon so the observer sees different shadow sizes of a single black hole at different polarized lights. Moreover, a phenomenological coupling between a photon and a generic vector field is introduced to analyze black hole shadows \cite{epb}. Notably, black hole shadows in edge-on view have different appearances for different frequencies of the observed light. The effects of the axion-photon interaction on the polarization patterns in black hole images have also been used to constrain axion-like dark matter around black holes \cite{tomoch}.

The Weyl tensor is an important tensor in general relativity because it describes a type of gravitational distortion in spacetime. Thus, electrodynamics with the Weyl corrections has been used to analyze the optical properties of curved spacetimes. In modified electrodynamics, the interaction between the Maxwell field and the Weyl tensor has a simple form. Specifically, the Weyl corrections could emerge naturally in quantum electrodynamics with the effective action of the photon that originates from one-loop vacuum polarization \cite{weyl0}. Although the Weyl corrections first emerged as an effective description of quantum effects, the extended theoretical models without the small coupling constant limit have been investigated for many physical motivations \cite{weyl1,weyl2,weyl3,weyl4,weyl5}. Recently, we have examined the effects of Weyl corrections on the shadow of a static phantom black hole \cite{weyl6} and find that a single black hole can have double shadows because natural light can be separated into two kinds of linearly polarized light beams along different propagation paths. We also probe the effects of Weyl corrections on the black hole image and its polarization distribution for a Schwarzschild black hole surrounded by an equatorial thin accretion disk \cite{weyl7}. The effects of Weyl corrections on strong gravitational lensing have been investigated in the Schwarzschild black hole spacetime \cite{weyl8} and the Kerr black hole spacetime \cite{weyl9}. However, in Ref.\cite{weyl9}, we have considered only a simple case where photons are limited in the equatorial plane \cite{weyl9}. The main reason is that, in the rotating black hole spacetime, the solutions describing the polarized photon motions are difficult to identify in the general case. Thus, how Weyl corrections affect the shadow of a rotating black hole is still an open issue. In this study, we utilize the small coupling approximation based on physical justification and obtain two solutions of polarized photon motions. One corresponds to ordinary light rays with the same propagation paths as in the case without Weyl corrections. The other corresponds to extraordinary light rays whose propagation paths depend on the Weyl corrections. With such special properties of extraordinary light rays, we can assess the effects of Weyl corrections on the shadow of a rotating black hole.

The paper is organized as follows: In Section II, we briefly introduce the equation of motion for the photons coupled to the Weyl tensor in the Kerr black hole spacetime and present two solutions for polarized photon motions in the small coupling approximation. In Section III, we numerically present the Kerr black hole shadows cast by extraordinary light rays with the Weyl corrections and probe their effects on the shadow. Finally, we end the paper with a summary.

\section{Equation of motion for photons coupled to a Weyl tensor in a Kerr black hole spacetime}
We first briefly review the equations of motion for photons coupled to a Weyl tensor in a Kerr black hole spacetime by the geometric optics approximation \cite{weyl0,Daniels,Caip,Cho1,Lorenci}. In the curved spacetime, the action contained in the electromagnetic field with Weyl corrections can be expressed as follows:
\begin{eqnarray}
S=\int {\rm d}^4x \sqrt{-g}\bigg[\frac{R}{16\pi
G}-\frac{1}{4}\bigg(F_{\mu\nu}F^{\mu\nu}-4\tilde{\alpha}
C^{\mu\nu\rho\sigma}F_{\mu\nu}F_{\rho\sigma}\bigg)\bigg],\label{acts}
\end{eqnarray}
where $C_{\mu\nu\rho\sigma}$ is the Weyl tensor with the following form:
\begin{eqnarray}
C_{\mu\nu\rho\sigma}=R_{\mu\nu\rho\sigma}-(
g_{\mu[\rho}R_{\sigma]\nu}-g_{\nu[\rho}R_{\sigma]\mu})+\frac{1}{3}R
g_{\mu[\rho}g_{\sigma]\nu}.
\end{eqnarray}
Here, $g_{\mu\nu}$ is the metric of the background spacetime and the brackets around the indices denote the antisymmetric part. $R_{\mu\nu\rho\sigma}$, $R_{\mu\nu}$, and $R$ are the Riemannian curvature tensor, Ricci tensor, and Ricci scalar, respectively. $F_{\mu\nu}$ is the usual electromagnetic tensor, and the coupling constant $\tilde{\alpha}$ has a dimension of length-squared. The coupling term with the Weyl tensor modifies the Maxwell equation as follows:
\begin{eqnarray}
\nabla_{\mu}\bigg(F^{\mu\nu}-4\tilde{\alpha}
C^{\mu\nu\rho\sigma}F_{\rho\sigma}\bigg)=0.\label{WE}
\end{eqnarray}
From the corrected Maxwell equation (Eq. (\ref{WE})), we can derive the equations of motion for coupled photons by resorting to the geometric optics approximation where the wavelengths of photons $\lambda$ are assumed to be smaller than a typical curvature scale but larger than the electron Compton wavelength \cite{weyl0,Daniels,Caip,Cho1,Lorenci}. In this manner, the electromagnetic field strength expressed in Eq. (\ref{WE}) can be simplified as follows:
\begin{eqnarray}
F_{\mu\nu}=f_{\mu\nu}e^{i\theta},\label{ef1}
\end{eqnarray}
where $f_{\mu\nu}$ is a slowly varying amplitude, and $\theta$ is a rapidly varying phase, which indicates that the derivative term $f_{\mu\nu;\lambda}$ is not dominated; thus, it can be neglected. The wave vector $k_{\mu}=\partial_{\mu}\theta$ can be regarded as the usual photon momentum in quantum theories. Combined with the Bianchi identity
\begin{eqnarray}
\nabla_{\lambda} F_{\mu\nu}+\nabla_{\mu} F_{\nu\lambda}+\nabla_{\nu} F_{\lambda\mu}=0,
\end{eqnarray}
the form of the amplitude $f_{\mu\nu}$ can be expressed as follows:
\begin{eqnarray}
f_{\mu\nu}=k_{\mu}a_{\nu}-k_{\nu}a_{\mu},\label{ef2}
\end{eqnarray}
where $a_{\mu}$ is the polarization vector satisfying the condition that $k_{\mu}a^{\mu}=0$. Substituting Eqs. (\ref{ef1}) and (\ref{ef2}) into Eq. (\ref{WE}), we can obtain the equations of motion for photons with the Weyl corrections, as follows:
\begin{eqnarray}
k_{\mu}k^{\mu}a^{\nu}+8\tilde{\alpha}
C^{\mu\nu\rho\sigma}k_{\sigma}k_{\mu}a_{\rho}=0.\label{WE2}
\end{eqnarray}
Thus, the Weyl corrections change the propagation of coupled photons in the background spacetime.
Moreover, by varying the action (Eq. (\ref{acts})) with respect to the metric $g_{\mu\nu}$, we can obtain the modified Einstein equation, as follows:
\begin{eqnarray}\label{estgravfield}
G_{\mu\nu}+8\tilde{\alpha}\pi G \mathcal{C}_{\mu\nu}=8\pi G T_{\mu\nu},
\end{eqnarray}
where $G_{\mu\nu}$ is the usual Einstein tensor. The forms of $\mathcal{C}_{\mu\nu}$ and $T_{\mu\nu}$ can be expressed as follows:
\begin{eqnarray}\label{tensor1}
\mathcal{C}_{\mu\nu}&=&-\nabla_{\rho}\nabla_{\sigma}(F_{\mu\nu}F^{\rho\sigma}+2F_{\mu}^{\;\sigma}F^{\rho}_{\;\nu}-2g_{\mu\nu}F^{\sigma}_{\;\lambda}F^{\rho\lambda})
-4\nabla_{\mu}\nabla_{\lambda}(F^{\lambda}_{\;\sigma}F_{\nu}^{\;\sigma})+\nabla_{\lambda}\nabla^{\lambda}(F_{\mu\sigma}F_{\nu}^{\;\sigma}-\frac{1}{3}g_{\mu\nu}F^{\rho\sigma}F_{\rho\sigma})\nonumber\\
&&+\nabla_{\mu}\nabla_{\nu}(F^{\rho\sigma}F_{\rho\sigma})+\frac{2}{3}[R_{\mu\nu}F^{\rho\sigma}F_{\rho\sigma}+2RF_{\mu}^{\;\sigma}F_{\nu\sigma}
-6R_{\rho\sigma}F_{\mu}^{\;\rho}F_{\nu}^{\;\sigma}]+F_{\mu}^{\;\lambda}F^{\rho\sigma}R_{\nu\lambda\rho\sigma}, \nonumber\\
T_{\mu\nu}&=&F_{\mu}^{\;\rho}F_{\nu\rho}-\frac{1}{4}g_{\mu\nu}F^{\rho\sigma}F_{\rho\sigma}.
\end{eqnarray}
Under the geometric optics approximation, we assume that the amplitude $f_{\mu\nu}$ of the electromagnetic field is small, slowly varying, and real. Notably, the tensors $\mathcal{C}^{\mu\nu}$ and $T^{\mu\nu}$ in the Einstein equation (Eq. (\ref{estgravfield})) are second-order small quantity $\mathcal{O}(f^2)$, but the Maxwell equation (Eq. (\ref{WE})) is the first-order equation of $f_{\mu\nu}$. Therefore, in the first-order approximation of $f_{\mu\nu}$, we find that the electromagnetic field and its interaction with the Weyl tensor do not correct the background spacetime. Thus, the electromagnetic field can be treated as a perturbational field. Based on these assumptions and approximations, the usual Kerr metric can be considered a solution to the action expressed in Eq. (\ref{acts}).
For a Kerr black hole, its metric has the following form with the standard Boyer-Lindquist coordinates:
\begin{eqnarray}
{\rm d}s^2&=&-\rho^2\frac{\Delta}{\Sigma^2}{\rm d}t^2+\frac{\rho^2}{\Delta}{\rm d}r^2+\rho^2
{\rm d}\theta^2+\frac{\Sigma^2}{\rho^2}\sin^2{\theta}({\rm d}\phi-\omega {\rm d}t)^2,\label{m1}
\end{eqnarray}
with
\begin{eqnarray}
\omega&=&\frac{2aMr}{\Sigma^2},
\;\;\;\;\;\;\;\;\;\;\;\;\;\;\;\;\;\;\;\;\;\;\;\;\;\;\;\;\;\rho^2=r^2+a^2\cos^2\theta,
\nonumber\\
\Delta&=&r^2-2Mr+a^2,\;\;\;\;\;\;\;\;\;\;\;\;\;\;\;\;\Sigma^2=(r^2+a^2)^2-a^2\sin^2\theta \Delta.
\end{eqnarray}
Here, the parameters $M$ and $a$ denote the mass and the spin parameter of the black hole, respectively. To build a local set of orthonormal frames in the Kerr black hole spacetime, we can resort to the vierbein fields defined as follows:
\begin{eqnarray}
g_{\mu\nu}=\eta_{ab}e^a_{\mu}e^b_{\nu},
\end{eqnarray}
where $\eta_{ab}$ is the Minkowski metric. The vierbein field $e^a_{\mu}$ has the following form:
\begin{eqnarray}
e^a_{\mu}=\left(\begin{array}{cccc}
\rho\frac{\sqrt{\Delta}}{\Sigma}&0&0&-\frac{\omega\Sigma}{\rho}\sin\theta\\
0&\frac{\rho}{\sqrt{\Delta}}&0&0\\
0&0&\rho&0\\
0&0&0&\frac{\Sigma}{\rho}\sin\theta
\end{array}\right),
\end{eqnarray}
and its inverse $e_a^{\mu}$ has the following form:
\begin{eqnarray}
e_a^{\mu}=\left(\begin{array}{cccc}
\frac{\Sigma}{\rho\sqrt{\Delta}}&0&0&0\\
0&\frac{\sqrt{\Delta}}{\rho}&0&0\\
0&0&\frac{1}{\rho}&0\\
\frac{\omega\Sigma}{\rho\sqrt{\Delta}}&0&0&\frac{\rho}{\Sigma\sin\theta}
\end{array}\right).
\end{eqnarray}
With the following notation for the antisymmetric combination of vierbein fields \cite{weyl0,Daniels}:
\begin{eqnarray}
U^{ab}_{\mu\nu}=e^a_{\mu}e^b_{\nu}-e^a_{\nu}e^b_{\mu},
\end{eqnarray}
the complete Weyl tensor in the Kerr black hole spacetime can have a simple form expressed as follows \cite{Daniels}:
\begin{eqnarray}
C_{\mu\nu\rho\sigma}&=&2\mathcal{A}\bigg(U^{01}_{\mu\nu}U^{01}_{\rho\sigma}
-U^{23}_{\mu\nu}U^{23}_{\rho\sigma}-U^{03}_{\mu\nu}U^{03}_{\rho\sigma}
+U^{12}_{\mu\nu}U^{12}_{\rho\sigma}\bigg)+2\mathcal{B}\bigg(
U^{02}_{\mu\nu}U^{02}_{\rho\sigma}-U^{13}_{\mu\nu}U^{13}_{\rho\sigma}-U^{03}_{\mu\nu}U^{03}_{\rho\sigma}
+U^{12}_{\mu\nu}U^{12}_{\rho\sigma}\bigg)\nonumber\\
&+&\mathcal{C}\bigg(U^{01}_{\mu\nu}U^{23}_{\rho\sigma}+
U^{23}_{\mu\nu}U^{01}_{\rho\sigma}-U^{03}_{\mu\nu}U^{12}_{\rho\sigma}
-U^{12}_{\mu\nu}U^{03}_{\rho\sigma}\bigg)+\mathcal{D}\bigg(
-U^{02}_{\mu\nu}U^{13}_{\rho\sigma}-
U^{13}_{\mu\nu}U^{02}_{\rho\sigma}-U^{03}_{\mu\nu}U^{12}_{\rho\sigma}
-U^{12}_{\mu\nu}U^{03}_{\rho\sigma}\bigg)\nonumber\\
&+&\mathcal{E}\bigg(
U^{01}_{\mu\nu}U^{02}_{\rho\sigma}+
U^{02}_{\mu\nu}U^{01}_{\rho\sigma}+U^{13}_{\mu\nu}U^{23}_{\rho\sigma}
+U^{23}_{\mu\nu}U^{13}_{\rho\sigma}\bigg)+\mathcal{F}\bigg(
U^{01}_{\mu\nu}U^{13}_{\rho\sigma}+
U^{13}_{\mu\nu}U^{01}_{\rho\sigma}-U^{02}_{\mu\nu}U^{23}_{\rho\sigma}
-U^{23}_{\mu\nu}U^{02}_{\rho\sigma}\bigg),
\end{eqnarray}
with
\begin{eqnarray}
\mathcal{A}&=&\frac{Mr}{\rho^6\Sigma^2}(r^2-3a^2\cos^2\theta)[
2(r^2+a^2)^2+a^2\Delta\sin^2\theta],\nonumber\\
\mathcal{B}&=&-\frac{Mr}{\rho^6\Sigma^2}(r^2-3a^2\cos^2\theta)[
(r^2+a^2)^2+2a^2\Delta\sin^2\theta],
\nonumber\\
\mathcal{C}&=&-\frac{aM\cos\theta}{\rho^6\Sigma^2}(3r^2-a^2\cos^2\theta)[
2(r^2+a^2)^2+a^2\Delta\sin^2\theta],\nonumber\\
\mathcal{D}&=&\frac{aM\cos\theta}{\rho^6\Sigma^2}(3r^2-a^2\cos^2\theta)[
(r^2+a^2)^2+2a^2\Delta\sin^2\theta],
\nonumber\\
\mathcal{E}&=&-\frac{3a^2M\sqrt{\Delta}\cos\theta}{\rho^6\Sigma^2}(3r^2-a^2\cos^2\theta)
(r^2+a^2)\sin\theta,\nonumber\\
\mathcal{F}&=&\frac{3aMr\sqrt{\Delta}}{\rho^6\Sigma^2}(r^2-3a^2\cos^2\theta)
(r^2+a^2)\sin\theta.
\end{eqnarray}
To obtain the equations of motion for coupled photons in the Kerr black hole spacetime, we can introduce three linear combinations of momentum components as follows \cite{weyl0,Daniels}:
\begin{eqnarray}
l_{\nu}=k^{\mu}U^{01}_{\mu\nu},\;\;\;\;\;\;\;\;\;\;
n_{\nu}=k^{\mu}U^{02}_{\mu\nu},\;\;\;\;\;\;\;\;\;\;
r_{\nu}=k^{\mu}U^{03}_{\mu\nu},\label{pvector1}
\end{eqnarray}
together with the following dependent combinations:
\begin{eqnarray}
&&p_{\nu}=k^{\mu}U^{12}_{\mu\nu}=\frac{1}{e_t^0k^t}\bigg[e_r^1k^rn_{\nu}
-e_{\theta}^2k^{\theta}l_{\nu}\bigg],\nonumber\\
&&m_{\nu}=k^{\mu}U^{23}_{\mu\nu}=\frac{1}{e_t^0k^t}
\bigg[e_{\theta}^2k^{\theta}r_{\nu}-(e_{t}^3k^{t}+e_{\phi}^3k^{\phi})n_{\nu}\bigg],\nonumber\\
&&q_{\nu}=k^{\mu}U^{13}_{\mu\nu}=\frac{1}{e_t^0k^t}
\bigg[e_{r}^1k^{r}r_{\nu}-(e_{t}^3k^{t}+e_{\phi}^3k^{\phi})l_{\nu}\bigg].\label{vect33}
\end{eqnarray}
These polarization vectors are orthogonal to the wave vector $k_{\nu}$. Contracting Eq. (\ref{WE2}) with the vectors $l_{\nu}$, $n_{\nu}$, $r_{\nu}$ and utilizing the relationship expressed in Eq. (\ref{vect33}), the equations of motion of photons with the Weyl corrections (Eq. (\ref{WE2})) can be further simplified as a set of equations for three independent polarization components, i.e., $a\cdot l$, $a\cdot n$, and $a\cdot r$:
\begin{eqnarray}
\bigg(\begin{array}{ccc}
K_{11}&K_{12}&K_{13}\\
K_{21}&K_{22}&
K_{23}\\
K_{31}&K_{32}&K_{33}
\end{array}\bigg)
\bigg(\begin{array}{c}
a \cdot l\\
a \cdot n
\\
a \cdot r
\end{array}\bigg)=0.\label{Kk1}
\end{eqnarray}
The coefficient $K_{ij}$ is complicated; thus, we do not list it here (for details on the coefficient $K_{ij}$, please refer to Eqs. (20)-(22) in Ref. \cite{weyl9}).
The necessary and sufficient condition for Eq. (\ref{Kk1}) to have nonzero solutions is that the determinant of its coefficient matrix is zero, i.e., $|K|=0$. However, in the Kerr black hole spacetime, finding a solution to satisfy $|K|=0$ in the general case is difficult because of the complicated coefficient $K_{ij}$. Here, we limit ourselves to the case where the coupling parameter $\tilde{\alpha}$ is small, which is physically justified. By retaining only the first-order term and ignoring other higher-order terms of $\tilde{\alpha}$, the determinant $|K|$ can be expanded with the following linear form:
\begin{eqnarray}
|K|=\mathcal{K}^2[\mathcal{K}-8\tilde{\alpha}(\mathcal{C}+\mathcal{D})e_r^1 k^r(e_t^3k^t+e_{\phi}^3k^{\phi})]+\mathcal{O}(\alpha^2),
\end{eqnarray}
with
\begin{eqnarray}
\mathcal{K}=[-(e_{t}^0)^2+(e_t^3)^2]k^tk^t+(e_r^1)^2 k^rk^r+(e_{\theta}^2)^2 k^{\theta}k^{\theta}+2e_{\phi}^3 e_t^3 k^t k^{\phi}+(e_{\phi}^3)^2 k^{\phi}k^{\phi}.
\end{eqnarray}
Thus, this small $\alpha$ approximation has two nonzero solutions for Eq. (\ref{Kk1}), i.e,
\begin{eqnarray}
&&[-(e_{t}^0)^2+(e_t^3)^2]k^tk^t+(e_r^1)^2 k^rk^r+(e_{\theta}^2)^2 k^{\theta}k^{\theta}+2e_{\phi}^3 e_t^3 k^t k^{\phi}+(e_{\phi}^3)^2 k^{\phi}k^{\phi}=0,\label{solution1}\\
&&[-(e_{t}^0)^2+(e_t^3)^2]k^tk^t+(e_r^1)^2 k^rk^r+(e_{\theta}^2)^2 k^{\theta}k^{\theta}+2e_{\phi}^3 e_t^3 k^t k^{\phi}+(e_{\phi}^3)^2 k^{\phi}k^{\phi}-8\tilde{\alpha}(\mathcal{C}+\mathcal{D})\nonumber\\&&\times e_r^1 k^r(e_t^3k^t+e_{\phi}^3k^{\phi})=0. \label{solution2}
\end{eqnarray}
The first solution (Eq. (\ref{solution1})) is independent of the coupling parameter $\tilde{\alpha}$, whose behavior is similar to those of ordinary light rays propagating in anisotropic crystals. The second solution (Eq. (\ref{solution2})) depends on the coupling parameter $\tilde{\alpha}$, whose behavior is similar to those of extraordinary light rays propagating in crystal optics. These findings indicate that the Weyl corrections split light rays in the Kerr spacetime into ordinary and extraordinary light rays and yield the birefringence phenomenon in spacetime. The curved paths of light propagations in a black hole spacetime means that the background spacetime can be equivalently regarded as a kind of nonhomogeneous media for light propagations. The coupling between the Weyl tensor and the electromagnetic field induces an extra stress-like effect on the media because the Weyl tensor is related to the tidal distortion due to gravity, which yields ordinary light and extraordinary light that have different refractivities and different propagation paths. In the subsequent section, we will further analyze the shadow caused by extraordinary light rays (Eq. (\ref{solution2})) and probe the corresponding effects of the Weyl corrections.
Eq. (\ref{solution2}) can be rewritten as follows:
\begin{eqnarray}
g^{\mu\nu}k_{\mu}k_{\nu}-\frac{8\tilde{\alpha}(\mathcal{C}+\mathcal{D})\sqrt{\Delta}}{\Sigma\sin\theta}k_{r}k_{\phi}=0.\label{solution22}
\end{eqnarray}
Because of the Weyl corrections, the wave vector $k_{\mu}$ of the coupled photon is not a null vector, and the propagation paths of extraordinary light rays are non-geodesic in the Kerr spacetime. However, as in Ref. \cite{Breton}, we can introduce an effective metric $\gamma_{\mu\nu}$ so that Eq. (\ref{solution22}) can be rewritten as follows:
\begin{eqnarray}
\gamma^{\mu\nu}k_{\mu}k_{\nu}\equiv g^{\mu\nu}k_{\mu}k_{\nu}-\frac{8\tilde{\alpha}(\mathcal{C}+\mathcal{D})\sqrt{\Delta}}{\Sigma\sin\theta}k_{r}k_{\phi} =0, \label{solution22}
\end{eqnarray}
which means that the wave vector $k_{\mu}$ is a null vector, and the coupled photon moves along the null geodesics in the effective spacetime with the metric $\gamma_{\mu\nu}$. Thus, the real trajectories of photons in electrodynamics with the Weyl corrections can be equivalently regarded as the non-geodesic paths in the original black hole spacetime or the null geodesic lines in the corresponding effective spacetime. The latter provides a convenient method to solve the real trajectories of photons. We also note that $k_{\mu}$ is still the momentum of the coupled photons along the non-null geodesics in real spacetime. The effective metric for the coupled photons propagating along extraordinary light rays in the Kerr black hole spacetime can be expressed as follows:
\begin{eqnarray}
{\rm d}s^2=\gamma_{tt}{\rm d}t^2+\gamma_{rr}{\rm d}r^2+\gamma_{\theta\theta}{\rm d}\theta^2+\gamma_{\phi\phi}{\rm d}\phi^2+2\gamma_{t\phi}{\rm d}t{\rm d}\phi+2\gamma_{tr}{\rm d}t{\rm d}r
+2\gamma_{r\phi}{\rm d}r{\rm d}\phi,\label{effmetric}
\end{eqnarray}
with
\begin{eqnarray}\label{PPMmetr}
&&\gamma_{tt}=-\bigg(1-\frac{2Mr}{\rho^2}\bigg),\quad\quad\quad \quad \gamma_{rr}=\frac{\rho^2}{\Delta}, \quad\quad\quad\quad \gamma_{\theta\theta}=\rho^2,\nonumber\\
&&\gamma_{\phi\phi}=\frac{\Sigma^2\sin^2\theta}{\rho^2}, \quad\quad\quad\quad \gamma_{t\phi}=-\frac{2Mar\sin^2\theta}{\rho^2},\nonumber\\
&&\gamma_{tr}=-\frac{8\alpha M^4a^2r\sin\theta\cos\theta(3r^2-a^2\cos^2\theta)}{\rho^6\Sigma\sqrt{\Delta}},\nonumber\\
&&\gamma_{r\phi}=\frac{4\alpha M^3 a\Sigma\sin\theta\cos\theta(3r^2-a^2\cos^2\theta)}{\rho^6\sqrt{\Delta}}.
\end{eqnarray}
Here, we rescale the coupling parameter $\tilde{\alpha}$ as $\alpha=\tilde{\alpha}/M^2$, which ensures that the parameter $\alpha$ is a dimensionless quantity. As $a=0$, we find that the effective metric reduces to that of the usual Schwarzschild black hole and does not depend on the coupling parameter $\alpha$. That is, in the first-order linear approximation of the determinant $|K|$ with respect to $\alpha$, the extraordinary light ray travels along the same path as the ordinary light ray; thus, the birefringence phenomenon does not occur in this case. Moreover, we note that the propagation of the rays limited in the equatorial plane is independent of the coupling because the coupling-dependent metric functions $\gamma_{tr}$ and $\gamma_{r\phi}$ vanish as $\theta=\frac{\pi}{2}$, which means that the photon rings do not depend on the coupling in such a small $\alpha$ coupling case. A similar situation also occurs in the case where the light rays propagate along the rotation axis ($ \theta=0,\;\pi$) of the black hole. Thus, in this small coupling approximation, the Weyl corrections do not yield the birefringence phenomenon if the light rays propagate along the rotation axis or in the equatorial plane of a rotating black hole.
The Hamiltonian of coupled photons moving along null geodesics in the effective spacetime (Eq. (\ref{PPMmetr})) can be expressed as follows:
\begin{eqnarray}
H(x,p)=\frac{1}{2}g^{\mu \nu}(x)p_{\mu}p_{\nu}=0.\label{hamiltonian}
\end{eqnarray}
Because the metric functions expressed in Eq. (\ref{PPMmetr}) are independent of the coordinates $t$ and $\phi$, there exist two conserved quantities, i.e., the energy of the photon $E_0$ and the $z$-component of the angular momentum $L_{z0}$. However, because of the existence of the $drdt$ and $drd\phi$ terms, the forms of $E_0$ and $L_{z0}$ are modified as follows:
\begin{eqnarray}
E_0=-p_{t}=-\gamma_{tt}\dot{t}-\gamma_{tr}\dot{r}-\gamma_{t\phi}\dot{\phi}, \quad L_{z0}=p_{\phi}=\gamma_{t\phi}\dot{t}+\gamma_{r\phi}\dot{r}+\gamma_{\phi\phi}\dot{\phi}.\label{conserved quantities}
\end{eqnarray}
With these two conserved quantities, we can obtain the following equations of null geodesics:
\begin{eqnarray}
\dot{t}&=&\frac{\gamma_{\phi\phi}E_0+\gamma_{t\phi}L_{z0}+(\gamma_{tr}\gamma_{\phi\phi}-\gamma_{t\phi}\gamma_{r\phi})\dot{r}}{
\gamma_{t\phi}^2-\gamma_{tt}\gamma_{\phi\phi}},\label{u1}\\
\dot{\phi}&=&\frac{\gamma_{t\phi}E_0+\gamma_{tt}L_{z0}+(\gamma_{tr}\gamma_{t\phi}-\gamma_{tt}\gamma_{r\phi})\dot{r}}{
\gamma_{tt}\gamma_{\phi\phi}-\gamma_{t\phi}^2},\label{u4}\\
\ddot{r}&=&\frac{\gamma_{t\phi}^2-\gamma_{tt}\gamma_{\phi\phi}}{\gamma_{rr}(\gamma_{t\phi}^2-\gamma_{tt}\gamma_{\phi\phi})
+\gamma_{r\phi}^2\gamma_{tt}+\gamma_{tr}^2\gamma_{\phi\phi}-2\gamma_{tr}\gamma_{t\phi}\gamma_{r\phi}}
\bigg[\frac{1}{2}\bigg(\gamma_{tt,r}\dot{t}^2-\gamma_{rr,r}\dot{r}^2+\gamma_{\theta\theta,r}\dot{\theta}^2+\gamma_{\phi\phi,r}\dot{\phi}^2
\nonumber\\ &+&2\gamma_{t\phi,r}\dot{t}\dot{\phi}-2\gamma_{tr,\theta}\dot{t}\dot{\theta}-2\gamma_{\theta\theta,\theta}\dot{r}\dot{\theta}
-2\gamma_{r\phi,\theta}\dot{\theta}\dot{\phi}\bigg)\nonumber\\
&+&\frac{\gamma_{t\phi}\gamma_{r\phi}-\gamma_{tr}\gamma_{\phi\phi}}{\gamma_{t\phi}^2-\gamma_{tt}\gamma_{\phi\phi}}
\bigg(\gamma_{tr,r}\dot{r}^2+\gamma_{tt,r}\dot{t}\dot{r}+\gamma_{tt,\theta}\dot{t}\dot{\theta}+\gamma_{tr,\theta}\dot{r}\dot{\theta}
+\gamma_{t\phi,r}\dot{r}\dot{\phi}+\gamma_{t\phi,\theta}\dot{\theta}\dot{\phi}\bigg)\nonumber\\
&+&\frac{\gamma_{tr}\gamma_{t\phi}-\gamma_{tt}\gamma_{r\phi}}{\gamma_{t\phi}^2-\gamma_{tt}\gamma_{\phi\phi}}
\bigg(\gamma_{r\phi,r}\dot{r}^2+\gamma_{r\phi,\theta}\dot{r}\dot{\theta}+\gamma_{t\phi,r}\dot{t}\dot{r}
+\gamma_{t\phi,\theta}\dot{t}\dot{\theta}+\gamma_{\phi\phi,r}\dot{r}\dot{\phi}+\gamma_{\phi\phi,\theta}\dot{\theta}\dot{\phi}\bigg)\bigg],
\label{uu2}\\
\ddot{\theta}&=&\frac{1}{2\gamma_{\theta\theta}}(\gamma_{tt,\theta}\dot{t}^2+2\gamma_{tr,\theta}\dot{t}\dot{r}+\gamma_{rr,\theta}\dot{r}^2-
2\gamma_{\theta\theta,r}\dot{r}\dot{\theta}
-\gamma_{\theta\theta,\theta}\dot{\theta}^2+\gamma_{\phi\phi,\theta}\dot{\phi}^2+2\gamma_{t\phi,\theta}\dot{t}\dot{\phi}+2\gamma_{r\phi,\theta}
\dot{r}\dot{\phi}).
\label{uu3}
\end{eqnarray}
The presence of the metric functions $\gamma_{tr}$ and $\gamma_{r\phi}$ yields the quantities $\dot{t}$ and $\dot{\phi}$ that depend on $\dot{r}$, which means that the motion of photons behaves differently from that of the non-coupling case. Thus, the Kerr black hole shadows cast by extraordinary light rays with the Weyl corrections are expected to exhibit a new behavior.

\section{Kerr black hole shadows cast by extraordinary light rays with the Weyl corrections}
Because the null equations (Eqs. (\ref{u1})-(\ref{uu3})) cannot be variable-separable, we have to use the ¡°backward ray-tracing¡± method \cite{sw,swo,astro,chaotic,binary,sha18,my,swo7,swo8,swo9,swo10} to numerically simulate the Kerr black hole shadows cast by extraordinary light rays with the Weyl corrections. In this method, the light rays are assumed to evolve from the backward-time observer, and the position of each pixel in the final image can be obtained by numerically solving the nonlinear differential equations (Eqs. (\ref{u1})-(\ref{uu3})). The shadow of the black hole in the observer¡¯s sky is determined by the pixels related to the light rays falling into the black hole.
Notably, the observer staying in the background spacetime does not depend on whether the photon couples to the Weyl tensor or not. Therefore, we must expand the local basis of observer $\{e_{\hat{t}},e_{\hat{r}},e_{\hat{\theta}},e_{\hat{\phi}}\}$ as a form in the coordinate basis $\{ \partial_t,\partial_r,\partial_{\theta},\partial_{\phi} \}$ of the original background spacetime rather than the effective spacetime, i.e.,
\begin{eqnarray}
\label{zbbh}
e_{\hat{\mu}}=e^{\nu}_{\hat{\mu}} \partial_{\nu}.
\end{eqnarray}
In other words, the transformation matrix $e^{\nu}_{\hat{\mu}}$ satisfies $g_{\mu\nu}e^{\mu}_{\hat{\alpha}}e^{\nu}_{\hat{\beta}}=\eta_{\hat{\alpha}\hat{\beta}}$ rather than $\gamma_{\mu\nu}e^{\mu}_{\hat{\alpha}}e^{\nu}_{\hat{\beta}}=\eta_{\hat{\alpha}\hat{\beta}}$. The metric $\eta_{\hat{\alpha}\hat{\beta}}$ is the usual Minkowski metric. In the Kerr spacetime (Eq. (\ref{m1})), we can conveniently derive the following decomposition \cite{sw,swo,astro,chaotic,binary,sha18,my,swo7,swo8,swo9,swo10,swo11}:
\begin{eqnarray}
\label{zbbh1}
e^{\nu}_{\hat{\mu}}=\left(\begin{array}{cccc}
\zeta&0&0&\gamma\\
0&A^r&0&0\\
0&0&A^{\theta}&0\\
0&0&0&A^{\phi}
\end{array}\right),
\end{eqnarray}
where $\zeta$, $\gamma$, $A^r$, $A^{\theta}$, and $A^{\phi}$ are real coefficients with the following forms:
\begin{eqnarray}
\label{xs}
&&A^r=\frac{1}{\sqrt{g_{rr}}},\;\;\;\;\;\;\;\;\;\;\;
A^{\theta}=\frac{1}{\sqrt{g_{\theta\theta}}},\;\;\;\;\;\;\;\;\;\;\;
A^{\phi}=\frac{1}{\sqrt{g_{\phi\phi}}},\;\;\;\;\;\;\;\;\;\;\;
\zeta=\sqrt{\frac{g_{\phi\phi}}{g^2_{t\phi}-g_{tt}g_{\phi\phi}}}.
\end{eqnarray}
Based on Eq. (\ref{zbbh}), the locally measured four-momentum $p^{\hat{\mu}}$ of a photon can be expressed as follows:
\begin{eqnarray}
\label{dl}
p^{\hat{t}}=-p_{\hat{t}}=-e^{\nu}_{\hat{t}} p_{\nu},\;\;\;\;\;\;\;\;\;
\;\;\;\;\;\;\;\;\;\;\;p^{\hat{i}}=p_{\hat{i}}=e^{\nu}_{\hat{i}} p_{\nu}.
\end{eqnarray}
Thus, the locally measured four-momentum $p^{\hat{\mu}}$ of coupled photons in the original Kerr spacetime (Eq. (\ref{m1})) can be further written as follows:
\begin{eqnarray}
\label{kmbh}
p^{\hat{t}}=\zeta E_0-\gamma L_{z0},\;\;\;\;\;\;\;\;\;\;\;\;\;p^{\hat{\phi}}=\frac{1}{\sqrt{g_{\phi\phi}}}p_{\phi},\;\;\;\;\;\;\;\;\;\;\;\;
p^{\hat{\theta}}=\frac{1}{\sqrt{g_{\theta\theta}}}p_{\theta},
\;\;\;\;\;\;\;\;\;\;\;\;\;\;\;\;
p^{\hat{r}}=\frac{1}{\sqrt{g_{rr}}}.
\end{eqnarray}
As in Refs. \cite{sw,swo,astro,chaotic,binary,sha18,my,swo7,swo8,swo9,swo10,swo11}, we can obtain the celestial coordinates for the pixel corresponding to extraordinary light rays with the Weyl corrections, as follows:
\begin{eqnarray}
\label{xd1}
x&=&-r_{obs}\frac{p^{\hat{\phi}}}{p^{\hat{r}}}=-r_{obs}\sqrt{\frac{g_{rr}}{g_{\phi\phi}}}\frac{p_{\phi}}{p_{r}}
=-r_{obs}\sqrt{\frac{g_{rr}}{g_{\phi\phi}}}\frac{\gamma_{t\phi}\dot{t}+\gamma_{r\phi}\dot{r}
+\gamma_{\phi\phi}\dot{\phi}}{\gamma_{tr}\dot{t}+\gamma_{rr}\dot{r}+\gamma_{r\phi}\dot{\phi}}, \nonumber\\
y&=&r_{obs}\frac{p^{\hat{\theta}}}{p^{\hat{r}}}=r_{obs}\sqrt{\frac{g_{rr}}{g_{\theta\theta}}}\frac{p_{\theta}}{p_{r}}=
r_{obs}\sqrt{\frac{g_{rr}}{g_{\theta\theta}}}\frac{\gamma_{\theta\theta}\dot{\theta}}{\gamma_{tr}\dot{t}+\gamma_{rr}\dot{r}+\gamma_{r\phi}\dot{\phi}},
\end{eqnarray}
where $r_{obs}$ and $\theta_{obs}$ are the radial coordinate and polar angle of the observer, respectively.

\begin{figure}
\centering
\includegraphics[width=4.6cm]{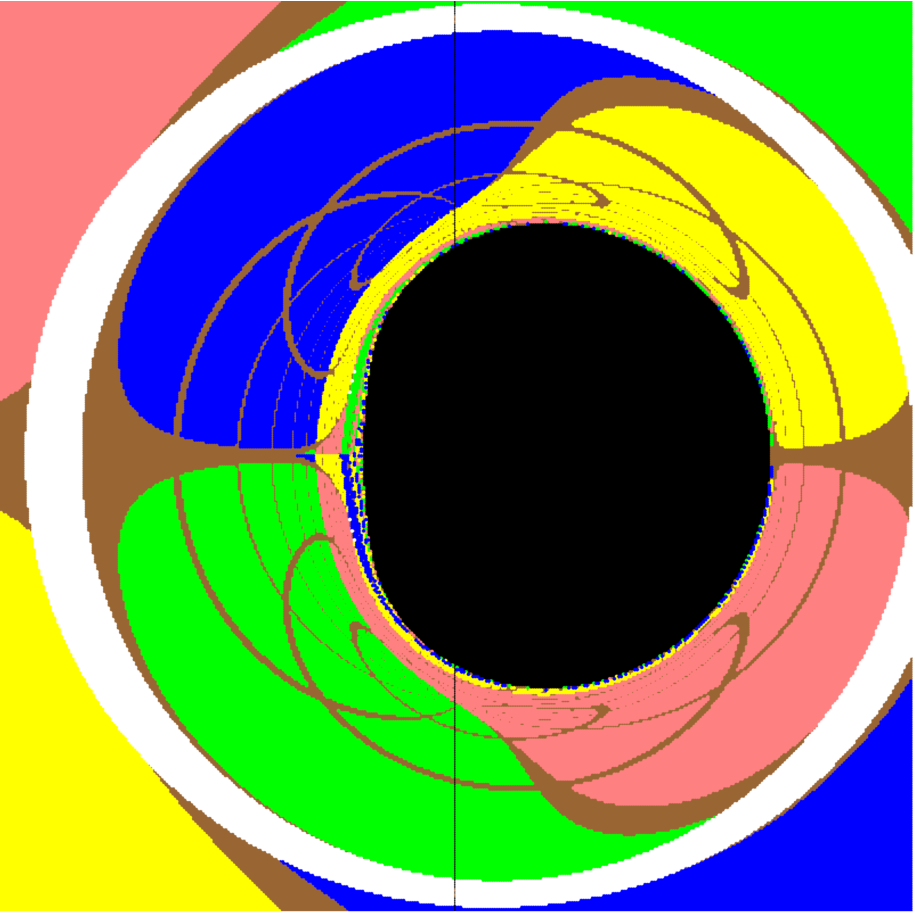}\;\;\includegraphics[width=4.6cm]{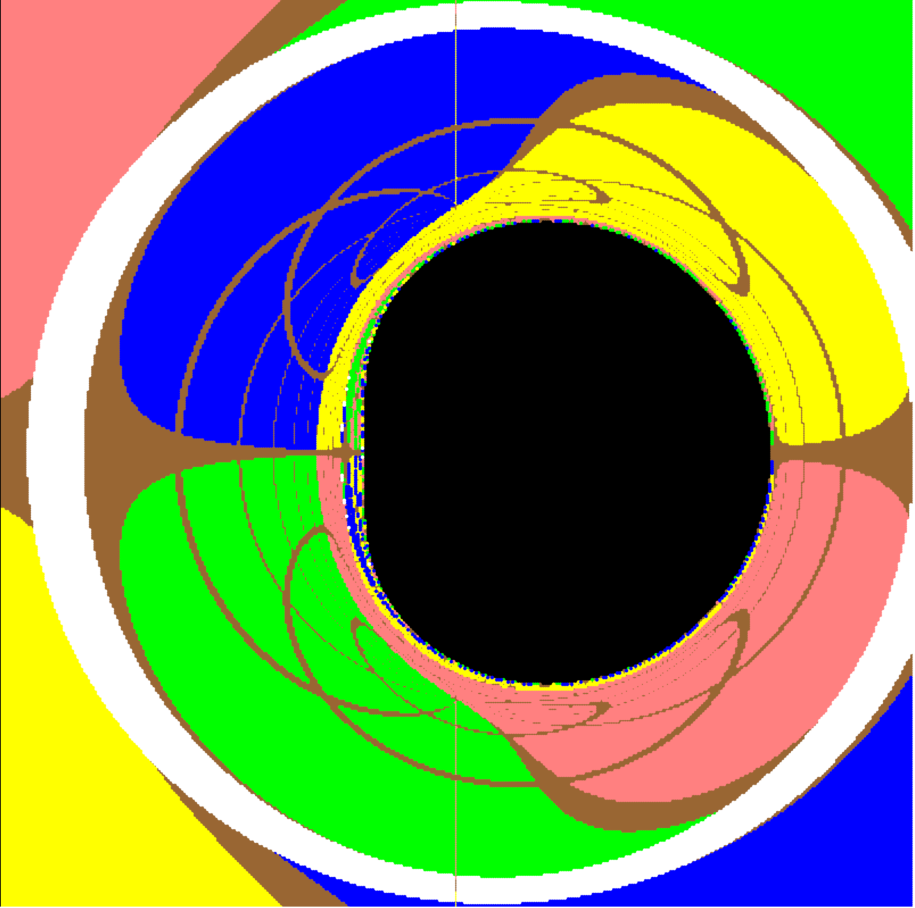}\;\;\includegraphics[width=4.6cm]{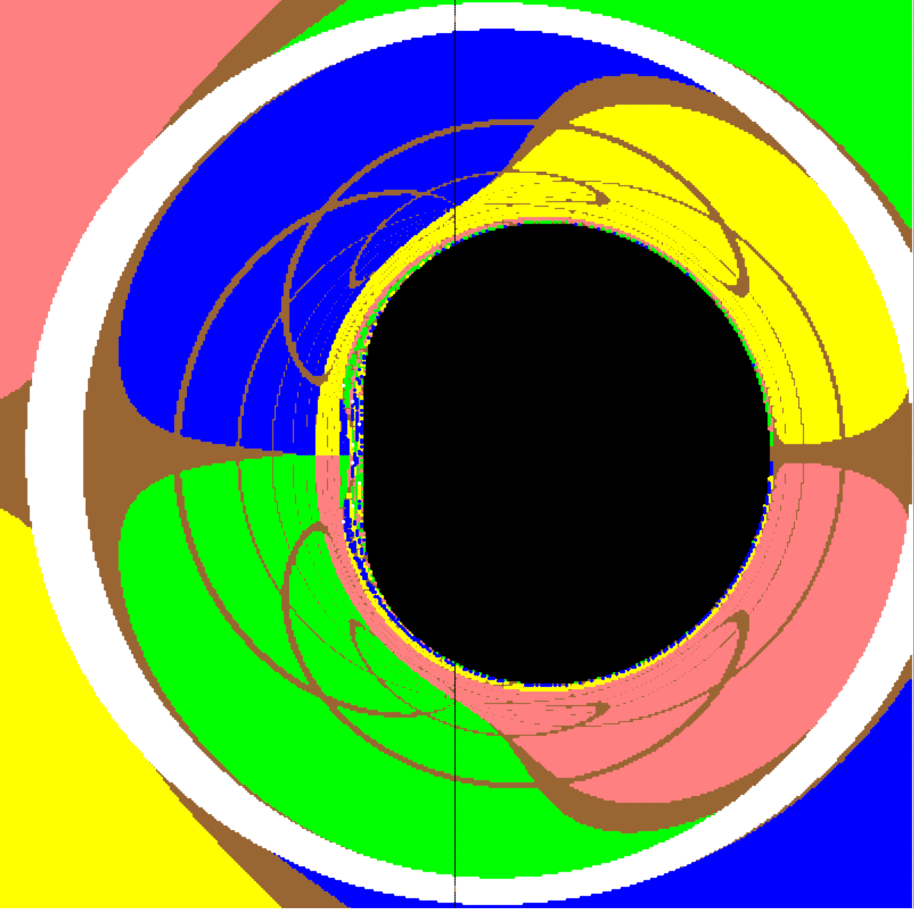}
\caption{Kerr black hole shadows cast by extraordinary light rays for different values of the coupling parameter $\alpha$ with fixed $a=0.998$. Here, we set the mass parameters $M=1$, $r_{obs}=30M$, and $\theta_{obs}=\pi/2$. The figures from left to right correspond to $\alpha=-0.3$, $0$, and $0.3$. }
\label{fig1}
\end{figure}

\begin{figure}
\centering
\includegraphics[width=5cm]{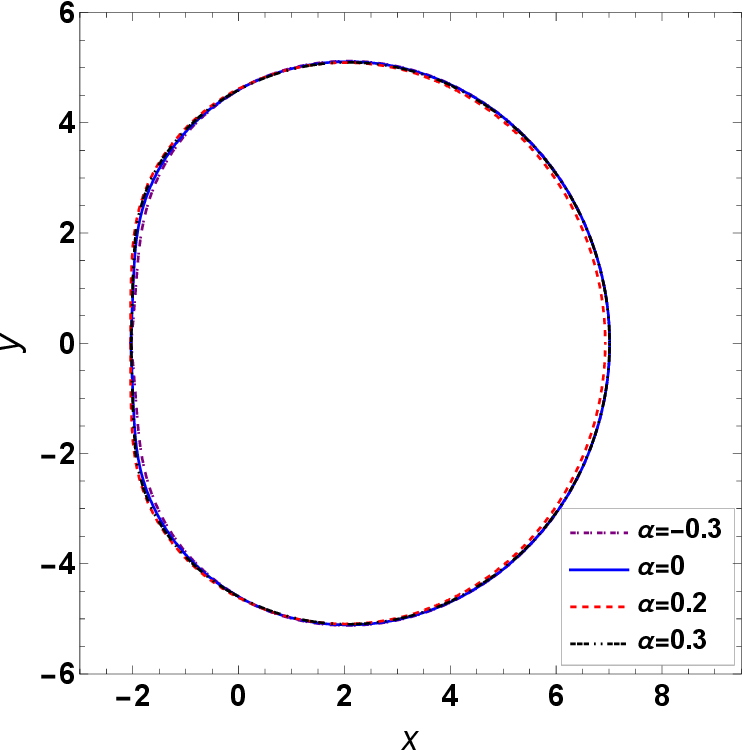}\;\;\includegraphics[width=5cm]{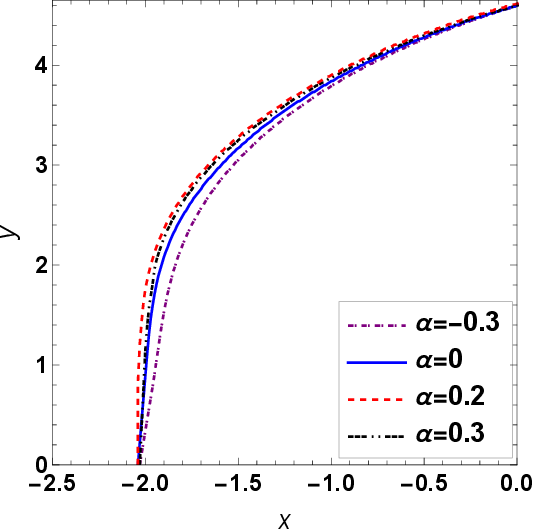}\;\;\includegraphics[width=5.1cm]{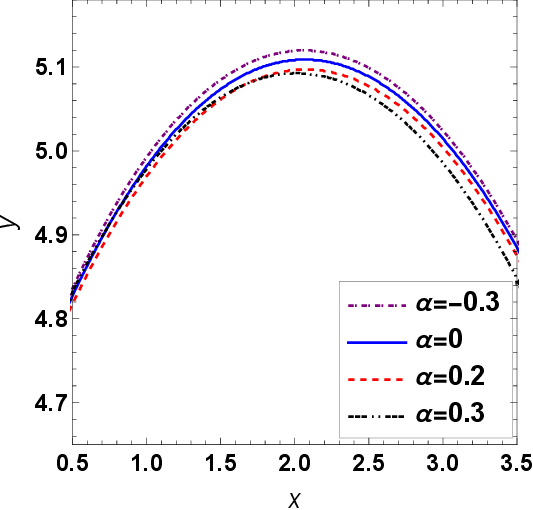}
\caption{Boundaries of Kerr black hole shadows cast by the extraordinary light rays arising from the coupling between the photon and the Weyl tensor. Here, we set $M=1$, $a=0.998$, $r_{obs}=30M$, and $\theta_{obs}=\pi/2$.}
\label{fig2}
\end{figure}

In Fig. \ref{fig1}, we present the Kerr black hole shadows cast by extraordinary light rays with the Weyl corrections as the observer lies in the equatorial plane. As in refs. \cite{sw,swo,astro,chaotic,binary,sha18,my,swo7,swo8,swo9,swo10}, the total celestial sphere is divided into four quadrants painted in different colors (i.e., green, blue, red, and yellow). The grids of the longitude and latitude lines are marked with adjacent brown lines separated by $10^\circ$. The distributions of the color regions and the shapes of the longitude and latitude lines in the figure reflect the bending of lights in the strong field regime near the black hole. The white ring in each figure, which is determined by lights from the reference spot lying in the line between the black hole and the observer, provides a direct demonstration of the Einstein ring. The black parts denote the black hole shadows. Here, we probe the effects of electrodynamics with the Weyl corrections on the shadows of the rapidly rotating black hole with $a=0.998$. The reason for selecting the black hole with a high spin parameter is that the corrections to the effective metric (Eq. (\ref{effmetric})) for extraordinary light rays in the Weyl corrected electrodynamics depend on the product of the parameters $\alpha$ and $a$, which indicates that the corrections are negligible in the slowly rotating case. Fig. \ref{fig1} shows that the positive coupling parameter $\alpha$ results in weak stretching in the vertical direction (i.e., $y$-axis direction) for the part of the black hole shadow with $x<0$ and weak squeezing along the $y$-axis direction for the part of the black hole shadow with $x>0$. The effects of the negative $\alpha$ on the black hole shadow are diametrically opposed. These properties are also further shown in Fig. \ref{fig2}. Moreover, for the positive coupling parameter case, we observe that the stretching effect on the part of the black hole shadow with $x<0$ increases with $\alpha$. However, with a further increase in $\alpha$, the corresponding stretching effect becomes weak. Similar behavior also manifests in the squeezing effects on the shadow part $x>0$ in this positive coupling parameter case.

\begin{figure}
\centering
\includegraphics[width=4.9cm]{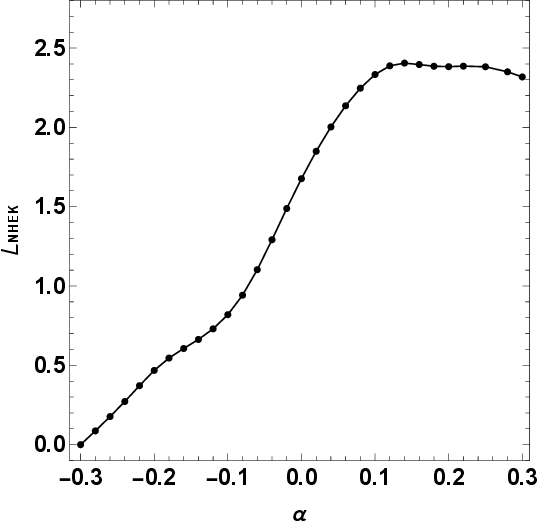}
\caption{Length of the near-horizontal extremal Kerr line in the black hole shadows cast by extraordinary light rays with the Weyl corrections. Here, we set $M=1$, $a=0.998$, $r_{obs}=30M$, and $\theta_{obs}=\pi/2$. }
\label{fig3}
\end{figure}

The well-known ¡°near-horizon extremal Kerr line (NHEK line)¡± is a vertical line segment on the edge of the shadow of a rapidly rotating Kerr black hole. The NHEK line is the brightest in the image of a rapidly rotating Kerr black hole, and its length is regarded to carry some characteristic information on the black hole \cite{lneh1,lneh2,lneh3,lneh4,lneh5}. In Fig. \ref{fig3}, we present the change in the length of the NHEK line with the Weyl coupling parameter $\alpha$ for the selected spin parameter $a=0.998$. The negative $\alpha$ decreases the length of the NHEK line, which can be attributed to its squeezing effects on the shadow for the part with $x<0$. Specifically, the length of the NHEK line becomes zero as $\alpha=-0.3$, which means the black hole shadow has a cusp shape in this case. For the positive $\alpha$, we observe that the length of the NHEK line first increases with $\alpha$, but with a further increase in $\alpha$, it nearly becomes a decreasing function of $\alpha$. The change in the length of the NHEK line with $\alpha$ is consistent with the stretching effect of $\alpha$ on the part of the black hole shadow with $x<0$. Finally, we do not detect a self-similar fractal structure in the black hole shadow. That is, in the small coupling case, the Weyl corrections are too weak to break the Kolmogorov-Arnold-Moser tori composed of photon orbits in the phase space. Therefore, no chaotic motion of coupled photons is detected, although their equations of motion (Eqs. (\ref{u1})-(\ref{uu3})) are generally not variable-separable.
In Table I, we compare the shadow size of the Kerr black hole with the shadow size of the Sgr A* and M87* black holes with different coupling values of $\alpha$. Here, we use the mass $M =4.0\times10^{6}M_{\odot}$, the observer distance $D_{O}=8.0kpc$, and the inclination angle $50^{\circ}$ for the black hole Sgr A* and the mass $M=6.5\times10^{9}M_{\odot}$, the observer distance $D_{O}=16.8Mpc$, and the inclination angle $17^{\circ}$ for the black hole M87*. The angular diameter $\theta_d$ does not depend on the coupling parameter $\alpha$ in the nonrotating black hole case but increases with $\alpha$ in the rotating black hole. The effect of $\alpha$ on the angular diameter $\theta_d$ increases with the black hole spin parameter. These properties of shadow are consistent with those in the previous discussion. The latest observation indicates that the angular diameters of the black holes M87* and Sgr A* are $\theta_d=42\pm3\mu as$ \cite{fbhs1} and $\theta_d=48.7\pm7\mu as$ \cite{fbhs1222}. By combining the data in Table I, we find that there is room for such a theoretical model of the electromagnetic field interacting with the Weyl tensor in the small coupling case.

\begin{table}[ht]\label{ts0}
\centering
\begin{tabular}{|c|c|c|c|c|c|c|c|c|c|c|c|c|c|c|c|}
\hline
\hline
\multicolumn{2}{|c|}{}&\multicolumn{7}{c|}{Sgr A*}&\multicolumn{7}{c|}{ M87 *}\\
\cline{3-16}
\multicolumn{2}{|c|}{\multirow{1}{*}{$\theta_{d}$ }}&\multicolumn{7}{c|}{$\alpha$ }&\multicolumn{7}{c|}{$\alpha$ }\\
\cline{3-16}\multicolumn{2}{|c|}{$(\mu as)$}&$-0.3$ &$-0.2$&$0.1$&$0$&$0.1$ &$0.2$&$0.3$&$-0.3$ &$-0.2$&$0.1$&$0$&$0.1$ &$0.2$&$0.3$\\
\hline
\multirow{3}{*}{$a$ }& 0.0 &51.190  & 51.190 &  51.190 &51.190  &51.190  & 51.190 & 51.190 &39.611  & 39.611 & 39.611 &39.611 &39.611  & 39.611& 39.611\\
\cline{2-16} & 0.5 &50.989 & 50.959 &50.928  &50.898   &50.868 & 50.838 & 50.798 &39.090 & 39.082 & 39.074 &39.067 &39.066  &39.059& 39.051     \\
\cline{2-16} & 0.998 &50.194 & 50.114 & 50.033 &$49.943$ &49.852 & 49.772 & $49.682$ &37.129 & 37.106 & 37.082 &$37.059$ &37.036 & 37.020 & $36.997$             \\
\hline\hline
\end{tabular}
\caption{Numerical estimation of the angular diameter of the black shadows of the supermassive black holes Sgr A* and M87* in the theoretical model of the electromagnetic field interacting with the Weyl tensor.}
\end{table}

\section{summary}
We have investigated the equation of motion for the photon coupled to the Weyl tensor in the Kerr black hole spacetime in the small coupling approximation. Our results show that the Weyl corrections yield the birefringence phenomenon in spacetime, where the propagation of ordinary light rays is independent of the coupling parameter $\alpha$, but the propagation of extraordinary light rays depends on the Weyl corrections. The effects of the Weyl corrections on extraordinary light rays depend on the product of the parameters $\alpha$ and $a$. Thus, the birefringence phenomenon does not occur in the nonrotating case because each extraordinary light ray travels along the same path as each ordinary light ray. Moreover, the birefringence phenomenon disappears as the propagation of the rays is limited in the equatorial plane or along the rotation axis of the black hole. We also probe the effects of the Weyl corrections on the black hole shadows cast by the extraordinary light rays in the rapidly rotating case. Notably, the positive coupling parameter $\alpha$ results in weak stretching (or squeezing) in the vertical direction (i.e., $y$-axis direction) for the part of the black hole shadow with $x<0$ (or $x>0$). The effects of the negative $\alpha$ on black hole shadow are diametrically opposed. In the case with a positive coupling parameter, we also find that the stretching effect on the part of the black hole shadow with $x<0$ first increases and then decreases with $\alpha$. Similar behavior also manifests in the squeezing effects on the shadow part with $x>0$. We also probe the change in the length of the NHEK line with the Weyl coupling parameter $\alpha$. The negative $\alpha$ decreases the length of the NHEK line, and the positive $\alpha$ first increases and then decreases the length of the NHEK line. The change in the length of the NHEK line with $\alpha$ is consistent with the stretching effect of $\alpha$ on the part of the black hole shadow with $x<0$. Finally, we observe no self-similar fractal structure in the black hole shadow in such a small coupling approximation.

\section{\bf Acknowledgments}
This work was  supported by the National Natural Science
Foundation of China under Grant No.12035005, 12275078, 11875026 and 2020YFC2201400.

\end{document}